\documentclass[aps,prb,twocolumn,showpacs,10pt,showkeys,preprintnumbers,amsmath,amssymb]{revtex4-1}
\usepackage{graphicx}% Include figure files
\usepackage{dcolumn}% Align table columns on decimal point
\usepackage{bm}% bold math
\usepackage{verbatim}
\usepackage[version=3]{mhchem}
\usepackage{subfigure}

\graphicspath{{images/}}

\begin{document}

\preprint{APS/123-QED}

\title{Dealloying of Platinum-Aluminum Thin Films \\ Part II. Electrode Performance}

%% use optional labels to link authors explicitly to addresses:
 \author{Thomas Ryll}
 \email{thomas.ryll@mat.ethz.ch}
 \author{Henning Galinski}
 \author{Lukas Schlagenhauf}
 \author{Felix Rechberger}
 \author{Sun Ying}
 \author{Ludwig J. Gauckler}
 \affiliation{Nonmetallic Inorganic Materials, ETH Zurich, Zurich, Switzerland}
 
 \author{Flavio C. F. Mornaghini}
 \author{Yasmina Ries}
 \author{Ralph Spolenak}
 \affiliation{Nanometallurgy, ETH Zurich, Zurich, Switzerland}
 
 \author{Max D\"obeli}
 \affiliation{Ion Beam Physics, ETH Zurich, Zurich, Switzerland}

\date{\today}

\begin{abstract}
% zu überarbeiten
Highly porous Pt/Al thin film electrodes on yttria stabilized zirconia electrolytes were prepared by dealloying of co-sputtered Pt/Al films. The oxygen reduction capability of the resulting electrodes was analyzed in a solid oxide fuel cell setup at elevated temperatures. During initial heating to 523~K exceptionally high performances compared to conventional Pt thin film electrodes were measured. This results from the high internal surface area and large three phase boundary length obtained by the dealloying process. Exposure to elevated temperatures of 673 K or 873 K gave rise to degradation of the electrode performance, which was primarily attributed to the oxidation of remaining Al in the thin films.
\end{abstract}
% PACS, the Physics and Astronomy
                             % Classification Scheme.
%Use showkeys class option if keyword
                              %display desired
\maketitle

%%
%% Start line numbering here if you want
%%
%\linenumbers

%% main text
\section{Introduction}
The efficiency of electrochemical devices like fuel cells is largely determined by the activity of the applied electrodes with respect to the oxygen reduction reaction (ORR). Up to date, Pt is the most common electrode material for polymer electrolyte membrane fuel cells (PEMFCs) \cite{Litster1, Erlebacher1} and low-temperature solid oxide fuel cells (LT-SOFCs). \cite{Su1, Evans1} 
%The ORR on Pt involves multiple reaction steps, namely oxygen adsorption, surface diffusion of intermediates, charge transfer and oxygen incorporation into the electrolyte, whose relative speed and spatial arrangement determine the overall reaction mechanism. Kinetic studies have shown that Pt electrodes operated below 1073 K are governed by surface diffusion. \cite{Mitterdorfer1} The situation is complicated, however, if one considers the strong interaction of Pt surfaces with oxygen. \cite{Ellinger1, Li1} While it is known that the oxidation of CO \cite{Ackermann1} or methane \cite{Hendriksen1} on Pt surfaces at atmospheric pressure is mainly due to surface oxides (compare Mars and Van Krevelen mechanism \cite{Doornkamp1}) the impact of oxidation and orientation of Pt surfaces on the ORR remain far from being understood. 
%
Attempts to increase the performance of Pt electrodes with respect to the ORR aim either at an increase of the inherent catalytic activity of Pt surfaces or the design of electrode morphologies that feature a higher density of catalytically active sites. Regarding the first approach, an optimization of the oxygen-metal interaction by alteration of the metal $d$ states either by alloying of Pt \cite{Stamenkovic1} or the fabrication of metal/Pt core-shell structures \cite{Mani1} have been shown to be promising. Guidelines for the second approach are a maximization of the electrode surface area and the three-phase-boundary (TPB) where the electrode is in contact with the ion conducting electrolyte and the gas phase. \cite{Litster1, Wang1} Furthermore, there is evidence that the grain boundaries of thin Pt layers provide oxygen pathways and contribute to the ORR, \cite{Ryll1} which is why small grains resulting in a high grain boundary density may be beneficial as well. 
An experimental procedure that is applied both in terms of the electronic modification of Pt surfaces and the fabrication of favorable Pt electrode morphologies is dealloying. \cite{Raney1} During dealloying the less noble element in a mostly binary alloy is selectively removed by an etchant. Depending on the experimental conditions, dealloying yields surfaces enriched in the more noble element, \cite{Koh1} nanowires \cite{Liu1} or nanoporous sponges. \cite{Erlebacher2} In the latter case, the resulting porosity is entirely interconnected and can exhibit pore sizes down to 3.4 nm in diameter and porosities up to 75 vol.\% as observed in the Pt/Cu system. \cite{Pugh1}
Over the last years, major research efforts have been targeted at more efficient catalysts for the ORR in PEMFCs. \cite{Litster1, Erlebacher1} In this context, dealloying has been used to produce core-shell nanoparticles and thin film surfaces enriched with Pt yielding a 2-6 fold increase of the electrochemical performance of Pt/Cu alloys compared to pure Pt. \cite{Mani1, Koh1, Yang1} Widely disregarded, however, has been the possibility to apply dealloyed electrodes catalyzing the ORR to SOFCs. %While nanoporous sponges are perfectly suited to catalyze the ORR from a morphological point of view, 
This originates in the apparent incompatibility of the inherent thermal instability of nanoporous metals with the traditionally high operating temperatures of SOFCs. Latter, however, have lately been decreased to values as low as 673 K, \cite{Su1} allowing to reconsider nanoporous Pt electrodes obtained by dealloying for SOFCs.  

%A well-known strategy towards high-performing Pt electrodes is a maximization of the three-phase-boundary (TPB) where the electrode is in contact with the ion conducting electrolyte and the gas phase. \cite{Litster1, Wang1} In terms of performance a high surface area is beneficial as well, given that surface diffusion plays a key role during the ORR on Pt electrodes. Finally, there is evidence that the grain boundaries of thin Pt layers provide oxygen pathways and contribute to the ORR. \cite{Ryll1} In summary, superior electrode activities are expected for Pt microstructures featuring a long TPB-length, a high surface area and a grain size in the nanometer range. A promising technique for the fabrication of nanoporous electrode layers meeting these requirements is dealloying. \cite{Raney1} During dealloying the less noble component of an alloy is selectively removed by an etchand leaving a nanoporous sponge of the more noble element behind. \cite{Erlebacher2} The resulting porosity is entirely interconnected and can exhibit pore sizes down to 3.4 nm in diameter as observed in the Pt/Cu system. \cite{Pugh1}\\ 
In this study, highly nanoporous Pt electrodes fabricated by dealloying of co-sputtered Pt/Al films on yttria stabilized zirconia (YSZ) are presented. We show morphological and electrochemical data that confirms the high catalytic activity of these layers and highlights their behavior at elevated temperatures. 
\section{Experimental}
Pt/Al films with a thickness of 370 nm were deposited on $\left\langle 100 \right\rangle$ oriented $\left(ZrO_2\right)_{0.905}(Y_2O_3)_{0.095}$ (YSZ) single crystals by magnetron co-sputtering at room temperature $(P_{Pt} = 37$ W$, P_{Al} = 252$ W$, P_{Ar} = 2.7*10^{-3}$ mbar$)$. The coated substrates were immersed for 10 min in a $4 M$ aqueous solution of NaOH resulting in dealloying of Al, formation of nanoporosity and film shrinkage. During dealloying the overall Al content decreased from 79 at.\% to 28 at.\% as determined by Rutherford backscattering spectrometry. Morphology, microstructure and electrochemical activity of dealloyed Pt/Al films were investigated before and after a heat treatment at 673 K for 10~h and 873 K for 14 h in air. Cross-sections were cut, polished and imaged using a NVISION 40 focused ion beam (FIB) system (Carl Zeiss NTS). Grazing incidence x-ray diffraction (GIXRD) of Cu K$_\alpha$ radiation was performed at a constant incidence angle of 1$^\circ$ using a X`Pert Pro MPD diffractometer (Panalytical). For the determination of lattice parameters and volume-weighted grain sizes, the resulting spectra were fitted by Pearson VII or Lorentzian distributions, respectively. During grain size analysis Scherrer`s formula using a shape factor of $0.9$ was applied. The instrumental line broadening was determined by means of a macroscopic Pt pellet. Dealloyed Pt/Al electrode films fabricated symmetrically on both sides of YSZ single crystals were electrochemically characterized by electrochemical impedance spectroscopy (EIS) % and contacted by sputtered Au frames. 
%Flattened Pt wires were attached to the Au surface by Au paste (C 5754 B, Heraeus) and ceramic glue (Feuerfestkitt, Firag). 
%Temperature was controled by a tube furnace constantly flushed with 50 ml/min of humidified air. During EIS,
using a Solartron 1260 impedance analyzer (Solartron Analytical) operated at an AC amplitude of 20~mV in the frequency range between 10 mHz and 4 MHz.  The real part of the low frequency impedance feature was multiplied by the electrode area in order to obtain the area specific electrode resistance (ASR).
\section{Results and Discussion}
\begin{figure}[t]
  \centering
    \subfigure[as-dealloyed]{\label{fig:dealloy1-a}\includegraphics[scale=0.2]{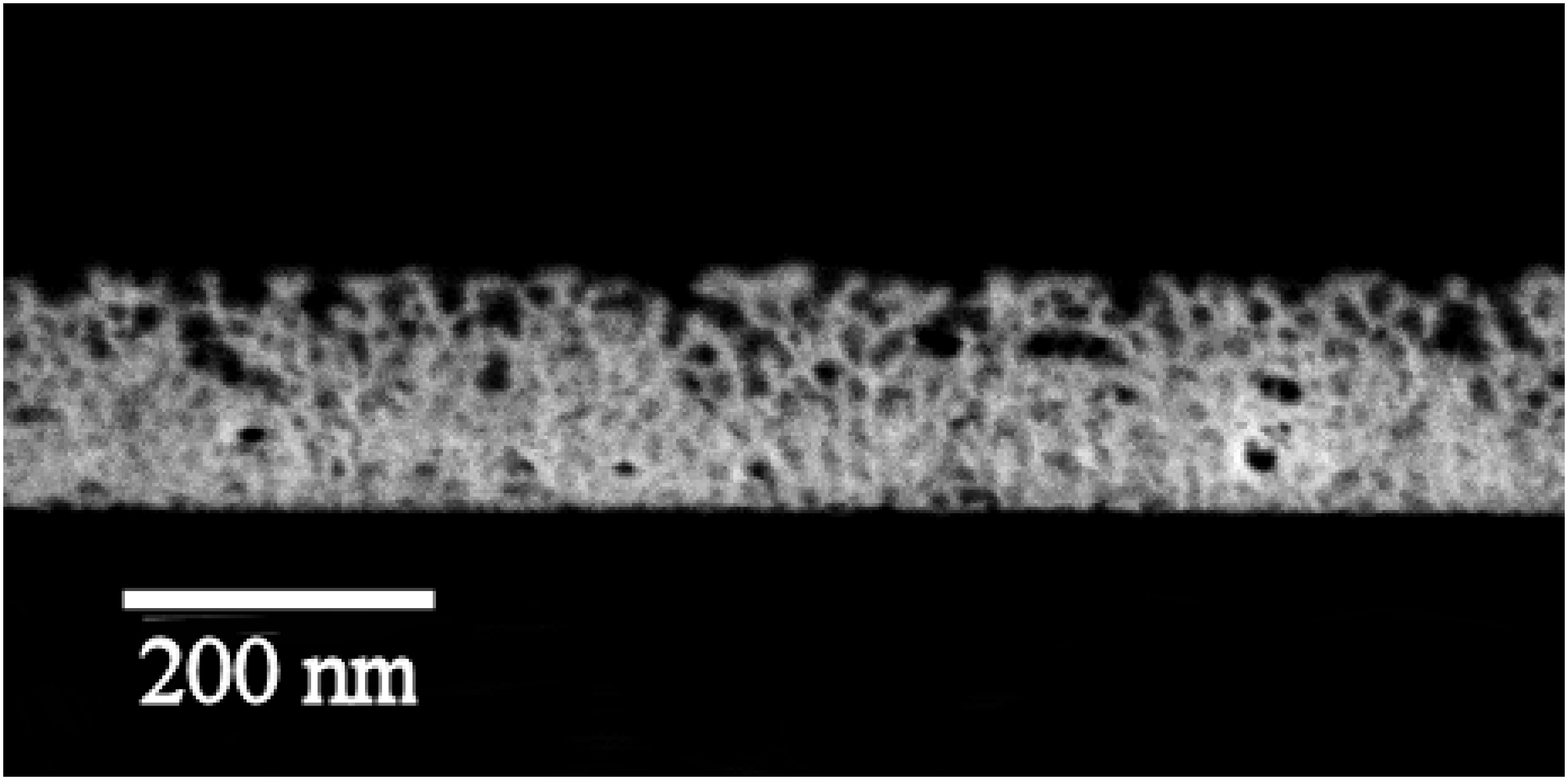}}
    \\
    \subfigure[after annealing]{\label{fig:dealloy1-b}\includegraphics[scale=0.1]{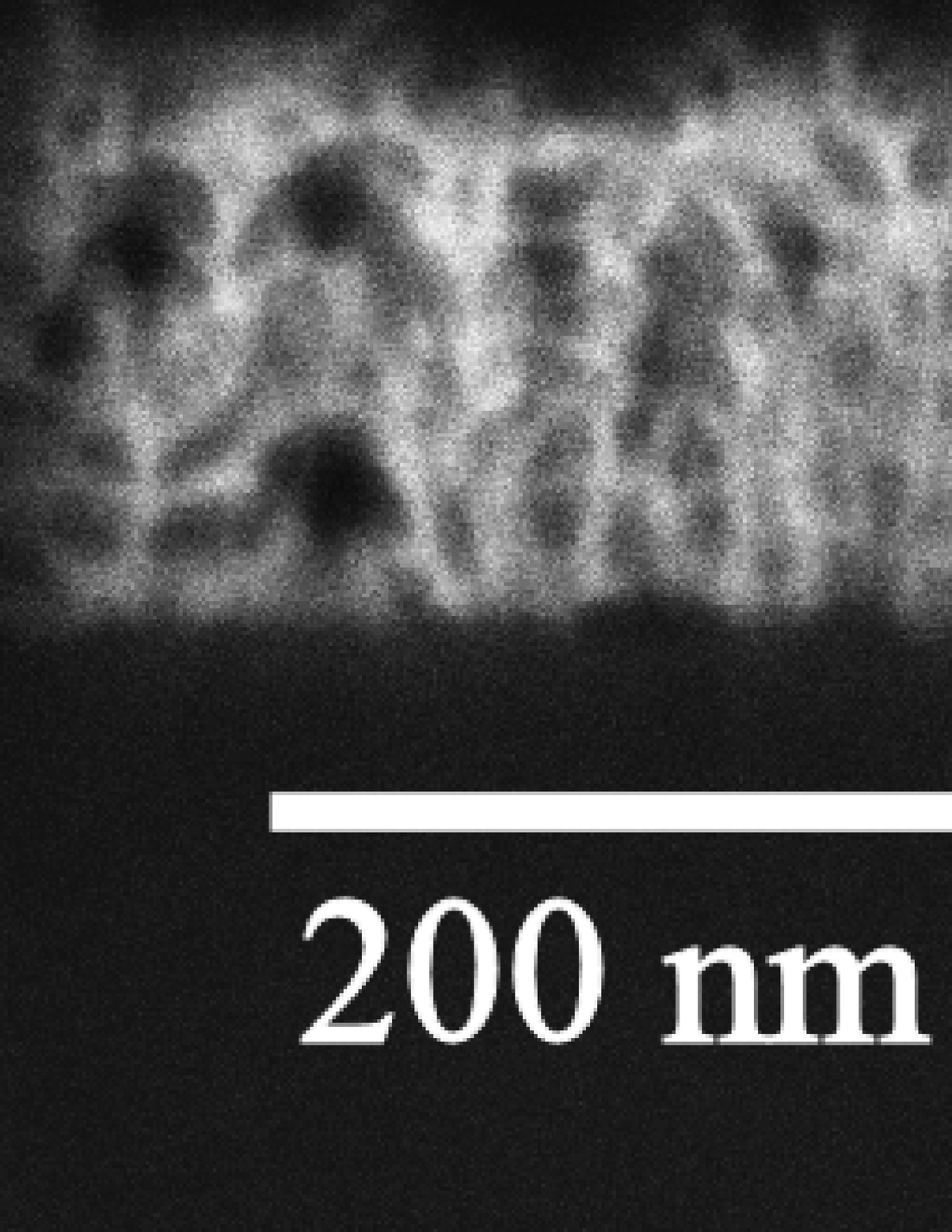}}
  \caption{	\ref{fig:dealloy1-a} FIB-polished cross-section of a nanoporous Pt$_{0.72}$Al$_{0.28}$ film on YSZ fabricated by dealloying. 
  					\ref{fig:dealloy1-b} Morphology after annealing at 673 K for 10 h and 873 K for 14 h in air.}
  					 
  \label{fig:dealloy1}
\end{figure}

The morphology of dealloyed Pt$_{0.72}$Al$_{0.28}$ films is shown in Figure \ref{fig:dealloy1-a}. The selective dissolution of Al from the Al-rich Pt$_{0.21}$Al$_{0.79}$ compound results in a branched nanoporosity with a mean pore intercept length of \mbox{$<L_2>$} = 10 nm. This value has to be considered a conservative estimate given that the smallest pores are not accessible by electron microscopic imaging of FIB-polished cross-sections. Since the stability of nanoporous electrode layers at elevated temperatures is a critical issue, dealloyed Pt/Al films were subjected to successive heating at 673 K for 10 h and 873 K for 14 h in air. The resulting morphology is shown in Figure \ref{fig:dealloy1-b}. The thermal treatment results in an increase of the mean pore intercept length \mbox{$<L_2>$} from 10 nm to 20 nm. This illustrates the high thermal stability of nanoporous Pt compared to nanoporous metals with a lower melting temperature. Nanoporous Pd, for example, has been reported to undergo coarsening from a ligament diameter of 15 nm to 180 nm during heating at 773 K already. \cite{Hakamada1} 

\begin{figure}[b]
	\centering
		\includegraphics[scale=0.3]{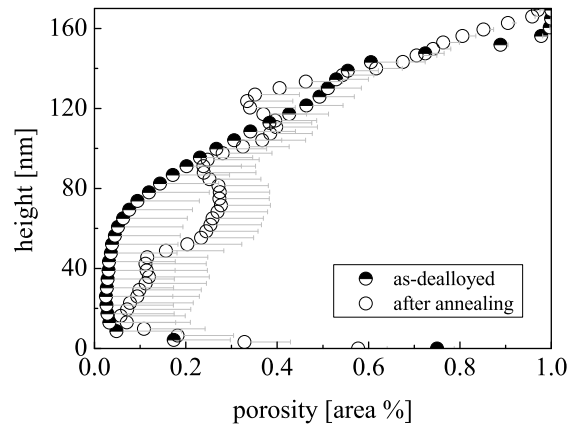}
		\caption{Porosity of Pt$_{0.72}$Al$_{0.28}$ films on YSZ plotted as a function of film height perpendicular to the film/substrate interface. Data for as-dealloyed films (partially filled circles) as well as films subjected to annealing at 673 K for 10 h and 873 K for 14 h in air (open circles) is shown.}
	\label{fig:dealloy2}
\end{figure}

\begin{figure}[t]
  \centering
    \includegraphics[scale=0.3]{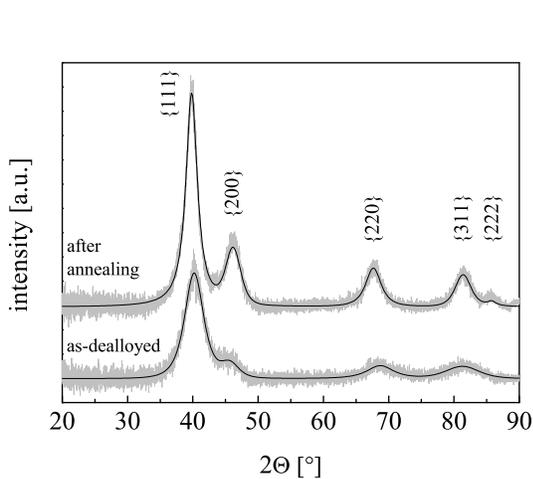}
  	\caption{GIXRD patterns of nanoporous Pt$_{0.72}$Al$_{0.28}$ films measured directly after dealloying or after additional annealing at 673 K for 10 h and 873 K for 14 h in air. Reflections are indexed as belonging to the structure group $Fm-3m$. The black line corresponds to a Pearson VII distribution fitted to the original data shown in grey.} 
  \label{fig:dealloy3}
\end{figure}

Both Figure \ref{fig:dealloy1-a} and \ref{fig:dealloy1-b} imply that the porosity is non-uniformly distributed over the film cross section and increases considerably towards the film surface. This becomes more obvious in Figure \ref{fig:dealloy2} where the porosity determined by means of multiple cross-sections of as-dealloyed and annealed Pt/Al films is plotted as a function of film height perpendicular to the film/substrate interface. 
As shown in Part I of this contribution, the porosity gradient results from the prolonged exposure of the surface-near film regions to the alkaline solution. \cite{Galinski1} The main difference between the morphology of as-dealloyed and annealed films appears in the thickness region between 20 nm and 90 nm where a higher porosity is found after annealing. This results from the preferred growth of small pores whose size is below the resolution limit of FIB imaging (roughly 5 nm) in the as-dealloyed state. The data shown in Figure \ref{fig:dealloy2} yields mean porosities of 34\% and 40\% before and after annealing, respectively.\\ 
In Figure \ref{fig:dealloy3}, GIXRD patterns of as-dealloyed and annealed Pt/Al films are compared. Five broad and partially overlapping reflections can be distinguished. The peak positions point at a face centered cubic lattice structure with space-symmetry $Fm-3m$ as exhibited by Pt as well as Al. The lattice parameters before and after annealing are shown in Table \ref{tab:microstructural parameters} and match the literature data for pure Pt. This implies that the remaining Al atoms are not homogeneously distributed in the Pt-rich film, but form a secondary phase whose volume fraction is too low to be detectable by GIXRD. As the reflections in both patterns appear in the expected intensity ratio no indication for texture is found. The considerable peak broadening traces back to a grain size $d$ that doubles from 2.4(9) nm to 5.0(9) nm during annealing. The microstructural parameters obtained by image analysis and GIXRD are summarized in Table \ref{tab:microstructural parameters}.\\ 
\begin{table}[b]
    \caption{\label{tab:microstructural parameters} Comparison of lattice parameter $a$, volume-weighted grain size $d$ and mean pore intercept length $<L_2>$ of nanoporous Pt$_{0.72}$Al$_{0.28}$ films before and after annealing at 673 K for 10 h and 873 K for 14 h in air. Literature values for the lattice parameter $a$ of Pt and Al are 0.39237 nm \cite{Owen1} and 0.40498 nm, \cite{Cooper1} respectively.}
\begin{ruledtabular}
\begin{tabular}{ccccc}
[nm]&as-dealloyed& after annealing\\
\hline
$a$ &0.391(5)&0.3923(8)\\
$d$ &2.4(9)&5.0(9)\\
$<L_2>$ &10&20\\
\end{tabular}
\end{ruledtabular}
\end{table}
\begin{figure}[t]
  \centering
    \subfigure[temperature dependence]{\label{fig:dealloy4-a}\includegraphics[scale=0.3]{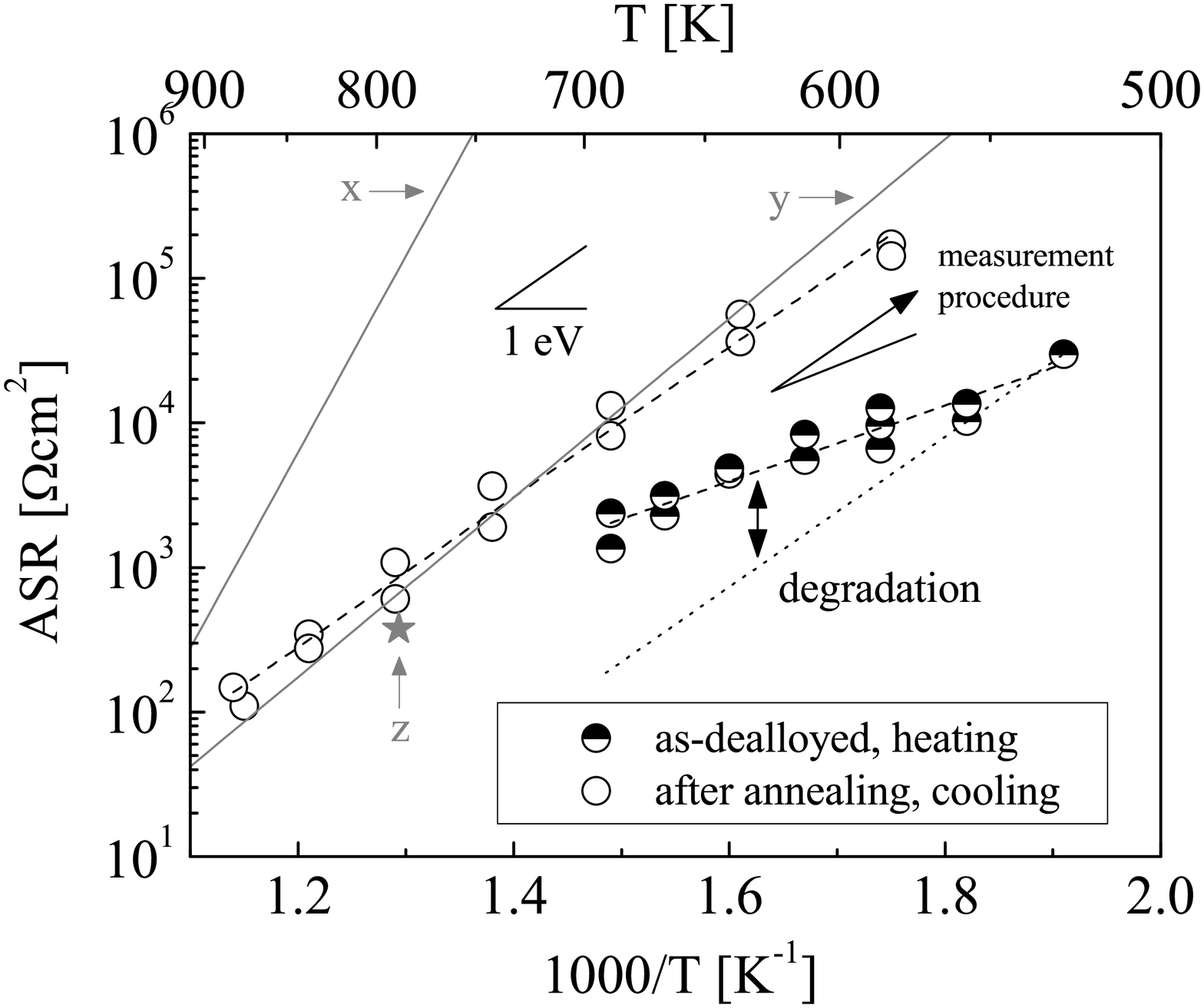}}
    \subfigure[time dependence]{\label{fig:dealloy4-b}\includegraphics[scale=0.3]{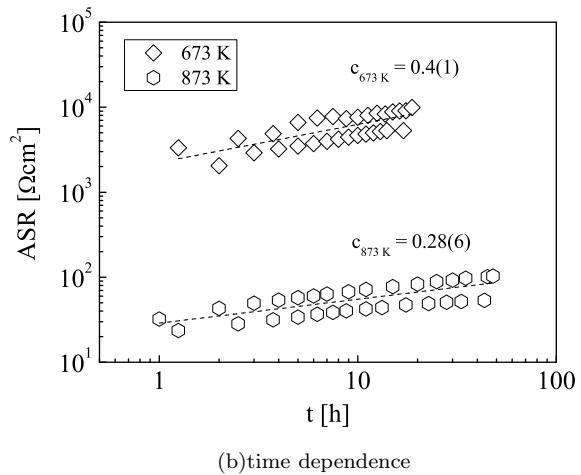}}
  \caption{ \ref{fig:dealloy4-a} Area specific resistance (ASR) of nanoporous Pt$_{0.72}$Al$_{0.28}$ films on YSZ plotted as a function of temperature T. Data of as-dealloyed films heated for the first time (partially filled circles) and films cooled after subsequent annealing at 673 K for 10 h and 873 K for 14 h in air (open circles) is compared. The dashed lines correspond to Arrhenius fits. Literature data for Pt thin film electrodes is shown in grey, x \cite{Verkerk1}, y \cite{Radhakrishnan1}, z \cite{Huang1}. 
  					\ref{fig:dealloy4-b} Temperature-induced increase of the ASR at 673 K (tilted squares) and 873 K (hexagons) plotted as a function of time t. The dashed line and the exponent c correspond to fitting of the data using a power law specified by equation (3).}
  					 
  \label{fig:dealloy4}
\end{figure}
\begin{table*}[t]
    \caption{\label{tab:electrochemical parameters} Experimentally determined parameters specifying the dependence of the ASR of nanoporous Pt$_{0.72}$Al$_{0.28}$ films on temperature T and time t. Whereas the parameters refering to temperature intervals correspond to Arrhenius fits given by the equations (1) and (2), the parameters refering to isothermal annealing at 673 K and 873 K correspond to the power law given by equation (3).}
\begin{ruledtabular}
\begin{tabular}{cccccc}
T [K]&ASR$_0$ [$\Omega$$cm^2$]&E$^{app}_{a}$ [eV]&$\Delta$$E_{deg}$ [eV]&b [$\Omega$$cm^2$]&c\\
\hline
$523-673$ (heating)&0.3(3)&0.52(5)&0.51(6)&-&-\\
$873-523$ (cooling)&1.7(9)$\times10^{-4}$&1.03(4)&0&-&-\\
$673$&-&-&-&2.2(6)$\times10^3$&0.4(1)\\
$873$&-&-&-&29(5)&0.28(6)\\
\end{tabular}
\end{ruledtabular}
\end{table*}
The area specific resistance (ASR) of nanoporous Pt/Al films with respect to the ORR is plotted in Figure \ref{fig:dealloy4-a} as a function of temperature. The data were measured during the first heating of as-dealloyed Pt films to 673~K as well as during cooling from 873 K after subsequent isothermal holds at 673 K and 873 K. All parameters obtained by fitting of the ASR data are summarized in Table \ref{tab:electrochemical parameters}. The initial ASR of dealloyed Pt/Al electrodes measured at 523 K is significantly lower than literature data for thin Pt electrodes at this temperature. This is attributed to the percolating nanoporosity of dealloyed Pt/Al electrodes resulting in a high internal surface area and large TPB-length. A temperature increase from 523~K to 673 K gives rise to an exponential decrease of the electrode resistance characterized by an apparent activation energy of $E^{app}_{a}$ = 0.52(5) eV. The fact that this energy is approximately half of the value expected for the ORR on thin Pt layers on the basis of literature (roughly 1 eV \cite{Wang1, Vladikova1}) points at a cumulative degradation of the electrode activity during heating according to

\begin{equation}
	ASR = ASR_{0}\times e^{\frac{E^{app}_{a}}{kT}} \text{}
\end{equation}\\
with
\begin{equation}
	E^{app}_{a} = E^{ORR}_{a} - \Delta E_{deg} \text{}
\end{equation}\\
where E$^{app}_{a}$ is the apparent activation energy measured during heating or cooling, $E^{ORR}_{a}$ is the activation energy of the ORR and $\Delta$$E_{deg}$ an activation drop due to degradation during heating. The ASR values measured after isothermal hold as well as the corresponding activation energy $E^{app}_{a}$ = 1.03(4) eV are in agreement with literature data. Presuming that no degradation occurs during cooling and, accordingly, $E^{app}_{a}$ = $E^{ORR}_{a}$ the activation energy of the degradation process is $\Delta$$E_{deg}$ = 0.51(6)~eV. At first, it seems obvious that the degradation of the electrode activity during heating is caused by thermally induced coarsening of the film morphology. The activation energy of the degradation process $\Delta$$E_{deg}$ = 0.51(6)~eV, however, is significantly smaller than the activation energies of self-diffusion processes on Pt that underlie coarsening (e.g. surface self-diffusion 0.69-0.84~eV, \cite{Bassett1} bulk self-diffusion 2.87 eV \cite{Schumacher1}). An alternative explanation for the observed degradation is surface segregation and adjacent oxidation of 
%either Pt or 
remaining Al during heating resulting in a gradual obstruction of electrochemically active sites. The oxidation of 
%bulk Pt or 
bulk Al is characterized by an activation energy in the range of 
%1.84 eV \cite{Fryburg1} and 
1.6-1.8 eV. \cite{Smeltzer1} In the case of particles in the nanometer range, however, the activation energy for oxidation can be reduced down to 0.33 eV as measured for Al nanopowders. \cite{Aumann1, Park1}
In order to gain additional insight into the electrode performance of nanoporous Pt/Al films, samples were electrochemically characterized during isothermal hold at 673 K and 873 K, respectively. The resulting development of the ASR with time is plotted in Figure \ref{fig:dealloy4-b}. The resistances increase continuously and can be fitted by a power law corresponding to  

\begin{equation}
	ASR = b\times t^{c} \text{}
\end{equation}\\
with exponents $c = 0.4(1)$ and $c = 0.28(6)$ at 673 K and 873 K, respectively. Accordingly, a nonlinear degradation behavior is found whose characteristic exponent is close to a square root time dependence at 673 K but decreases with increasing annealing temperature or time. The square root or, equivalently, parabolic time dependence of the electrochemical performance indicates rate determination by diffusion through a growing oxide layer (e.g. \cite{Bridges1}). This substantiates the hypothesis of electrode deactivation by oxidation of remaining Al. A transition towards a subparabolic oxidation characteristics at higher temperatures implies an impairment of diffusion and has been frequently observed during the oxidation of complex alloys (e.g. \cite{Barsoum1, Niranatlumpong1}). A reason for the subparabolic oxidation observed in this study might be a depletion of the oxidizing species resulting in a reduced growth rate of the oxide layer. \cite{Niranatlumpong1}
\section{Summary}
We demonstrate the fabrication of nanoporous Pt/Al films by selective dissolution of Al from co-sputtered Pt-Al alloy films. The dealloyed films exhibit a sponge-like morphology with a percolating nanoporosity in the range of 10 nm. Successive annealing at 673 K and 873 K results in minor coarsening only, which substantiates the high thermal stability of nanoporous Pt compared to other nanoporous metals. A favorable morphology in conjunction with a high thermal stability qualify these layers for electrode applications even at the elevated operating temperatures of SOFCs. The initial electrochemical performance of nanoporous Pt/Al films measured at 523 K exceeds literature data for this temperature. \cite{Verkerk1, Radhakrishnan1} With increasing temperature, however, electrochemical degradation occurs that is characterized by an activation energy of $\Delta$$E_{deg}$ = 0.51(6) eV and attributed to oxidation of remaining Al rather than coarsening of the nanoporous network. Future studies will focus on the mechanism of electrochemical degradation and the fabrication of nanoporous Pt/Al electrodes containing less Al. 
\begin{acknowledgments}
The authors greatfully acknowledge financial assistance by the Swiss Bundesamt f\"ur Energie (BfE), Swiss Electric Research (SER), the Competence Center Energy and Mobility (CCEM) and the Swiss National Foundation (SNF). Sincere thanks are given to the Electron Microscopy Center, ETH Zurich (EMEZ) for support. Thomas Ryll thanks Anna Evans for proofreading.   
\end{acknowledgments}

%% References with bibTeX database:
%\newpage %Just because of unusual number of tables stacked at end
\bibliographystyle{apsrev4-1}

\bibliography{BIB/LiteraturDealloying}% Produces the bibliography via BibTeX.

\end{document}